\documentclass[aps,preprint,onecolumn]{revtex4}
\usepackage{graphicx}

\begin{document}
\bibliographystyle{prsty}

\title{Near-field optical microscopy with a nanodiamond-based single photon tip }
\author{Aur\'{e}lien Cuche$^1$, Aur\'{e}lien Drezet$^1$, Yannick Sonnefraud $^{1,3}$, Orestis Faklaris$^{2}$, Fran\c{c}ois Treussart$^{2}$, Jean-Fran\c{c}ois Roch$^{2}$, Serge Huant$^{1}$}
\affiliation{$^1$~ Institut N\'{e}el, CNRS and Universit\'{e} Joseph Fourier, BP 166, 38042 Grenoble, France\\
 $^2$~Laboratoire de Photonique Quantique et Mol\'{e}culaire, UMR 8537 CNRS/Ecole Normale Sup\'{e}rieure de Cachan, Cachan, France\\
$^3$~New address: Experimental Solid State Physics, Blackett Laboratory, Imperial College London, Prince Consort Road, London SW7 2BZ, UK}

\email{serge.huant@grenoble.cnrs.fr} 
\date{\today}

\begin{abstract}
We introduce a point-like scanning single-photon source that operates at room temperature and offers an exceptional photostability (no blinking, no bleaching). This is obtained by grafting in a controlled way a diamond nanocrystal (size around 20 nm) with single nitrogen-vacancy color-center occupancy at the apex of an optical probe. As an application, we image metallic nanostructures in the near-field, thereby achieving a near-field scanning single-photon microscopy working at room temperature on the long term. Our work may be of importance to various emerging fields of nanoscience where an accurate positioning of a quantum emitter is required such as for example quantum plasmonics.
\end{abstract}
\maketitle
\section{Introduction}
An ideal source of light for various rapidly developing fields such as quantum-optics~\cite{1,2}, optomechanics~\cite{3,4} and plasmonics~\cite{5,6,7} would consist of a single quantum emitter with extreme photostability at room-temperature (RT) and adjustable position with nanometer accuracy in all three dimensions. In this respect, active tips made of a single fluorescent object attached to an optical tip are very promising since they can benefit from progresses made in tip manufacturing and nanopositioning systems for scanning-probe microscopy, e.g. near-field scanning optical microscopy (NSOM). Here, we introduce a point-like scanning single-photon source that operates at room temperature and offers an exceptional photostability (no blinking, no bleaching at all). This is obtained by grafting within an all-optical process a diamond nanocrystal (size around 20 nm) with single nitrogen-vacancy color-center occupancy at the apex of an optical probe. As an application, we image metallic nanostructures in the near-field, thereby achieving a stable RT near-field scanning single-photon microscopy. This microscopy opens up new possibilities for some emerging branches of nanoscience where an accurate positioning of a quantum emitter - or detector- is required such as for example quantum plasmonics~\cite{5,6,7} and high-resolution high-sensitivity magnetometry~\cite{8,9,10}.\\
Manufacturing a point-like scanning single-photon source faces a doubly challenging issue: a suitable quantum emitter must be identified and then attached onto the tip. In principle, single molecules with their point-like transition dipole moment~\cite{11,12} are ideally suited but in practice they are limited to low-temperature operation and eventually suffer from photobleaching. Colloidal semiconductor nanocrystals work in ambient conditions and are single-photon emitters~\cite{13}. However, despite promising progress in photostability~\cite{14}, they still undergo intermittency of their emission~\cite{15} (i.e. blinking) which can be sensitive at the single object level~\cite{16} (a notable exception are CdZnSe/ZnSe  nanocrystals~\cite{17} reported very recently) and can possibly bleach as well~\cite{17,18}. Interestingly, epitaxial quantum dots are very photostable but, despite the steady increase in their working temperature~\cite{19}, they still do not work at RT and are very difficult to manipulate in 3D. Rare-earth doped oxide nanoparticles of sizes in the sub-10 nm range are extremely photostable RT emitters~\cite{20,21} but, in spite of their small size, their fluorescence originates from a large ensemble of doping ions, so that single-photon emission remains elusive so far.\\
Color-centers in diamond~\cite{22}, in particular nitrogen-vacancy (NV) centers, appear to reconcile all of the above criteria since they are RT single-photon emitters~\cite{23,24}, their photostability is well-established~\cite{23} (no blinking, no bleaching) and they are hosted by nanocrystals with   steadily decreasing sizes thanks to progresses in materials processing~\cite{25,26,27,28}. Early use of NV-center doped diamond nanocrystals in NSOM active tips~\cite{29} was, however, limited by the size of the hosting crystal which was beyond the 50 nm range, so that the promise of single NV-occupancy, i.e. single-photon emission, was counterbalanced by size excess that prevents positioning with nanometer accuracy. The recent spectacular reduction in size~\cite{25,26,27,28} of fluorescent nanodiamonds (NDs), down to approximately 5 nm in the latest reports~\cite{28}, suggests that such limitation no longer exists and that active optical tips made of an ultra-small (well below 50 nm in size) ND with single NV-occupancy should be possible to achieve.\\

In this article, we describe a simple and thoughtfully easy-to-reproduce method for developing ND-based scanning single-photon sources and, in a proof-of-principle experiment, we validate such sources in NSOM imaging, thereby realizing what turns out to be the first scanning single-photon microscopy working at room temperature on a long term. We anticipate many interesting applications to this new optical microscopy.
\section{Principle and results}
\begin{figure}[h]
\begin{center}
\includegraphics[width=8cm]{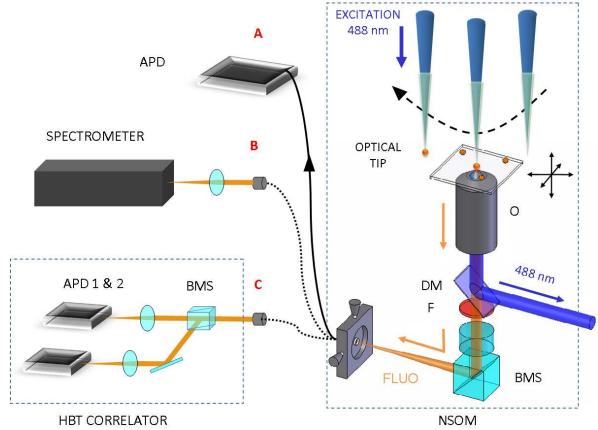}
\caption{Scheme of the optical setup used for tip functionalization with a single fluorescent nanodiamond (ND); (O= microscope objective, DM= dichroic mirror, F= optical filters, BMS= beamsplitter, APD= avalanche photodiode in the single-photon counting mode). The optical excitation is launched from the polymer-coated optical tip and the NV-center fluorescence is collected by a high NA objective, filtered, and injected into a multimode optical fiber. The latter can be connected either to an avalanche photodiode (channel A), a spectrometer (channel B), or a HBT correlator (channel C). This figure also depicts the protocol used for grafting a single ND at the tip apex: as soon as the position of the selected ND is reached during lateral scanning and concomitant monitoring of the fluorescence signal through channel A, the tip is temporarily approached vertically to the surface and lifted back to its original height (the dashed line is a scheme of the tip trajectory). Subsequent optical analysis checks that the trapped ND hosts a single NV center (photon correlation, channel C) and determines its charge state (fluorescence spectrum, channel B).}
\end{center}
\end{figure}
Our scanning single-photon sources are produced in a single transmission NSOM~\cite{18,26} environment. We successively use the optical tip for the imaging and selection of the very ND to be grafted at the tip apex, controlled attachment of the latter, and subsequent NSOM imaging of test surfaces. A sketch of the optical setup is shown in Fig. 1. The NV-center emission is excited with the 488 nm line of an Ar$^+$-Kr$^+$ continuous-wave laser that is injected into an uncoated (not metalized on its side) optical tip and is collected into a multimode optical fiber (core diameter: 50 $\mu$m) through a 60$\times$, NA 0.95 dry microscope objective. The remaining excitation light is removed by means of a dichroic mirror complemented either by a band-pass filter centered at 607$\pm35$ nm for photon counting and imaging or a long-pass filter ($>$ 532 nm) for spectra acquisition of the neutral and negatively-charged NV centers~\cite{30}. The collection fiber can be connected either to an avalanche photodiode detector (SPCM-AQR 16, Perkin-Elmer, Canada) for imaging and optical control of ND manipulation (channel A), to a charge-coupled device attached to a spectrometer (channel B), or to a Hanbury Brown and Twiss (HBT) correlator (channel C). In the HBT module, the signal is separated by a 50/50 beamsplitter and focused onto two avalanche photodiodes (SPCM-AQR 13, Perkin-Elmer, Canada) that are linked to a time-correlated single-photon counting module (PicoHarp 300, PicoQuant, Germany). A long-pass filter (750 nm) and a diaphragm placed in front of each detector eliminate most of the detrimental optical cross-talk.
\begin{figure}[p]
\begin{center}
\includegraphics[width=8cm]{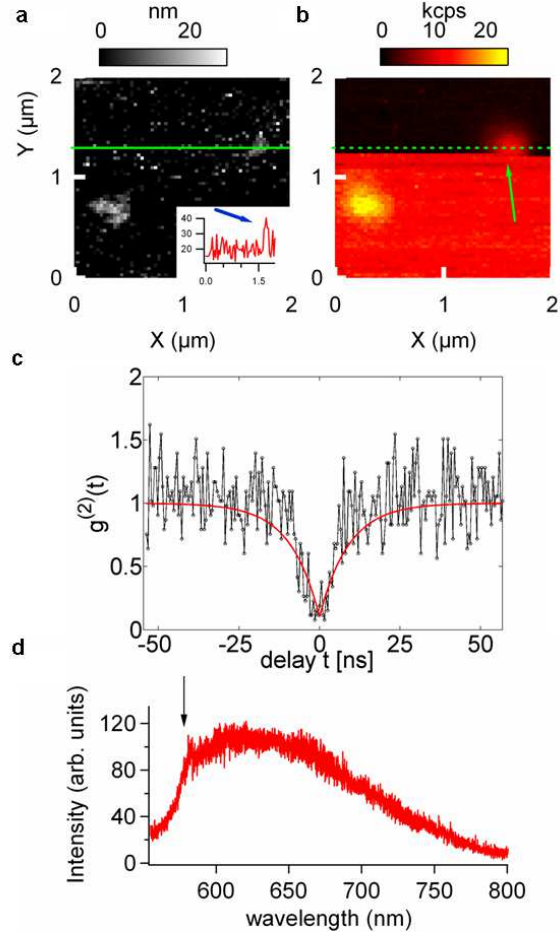}
\caption{(a) Numerically flattened topographic and (b) fluorescence NSOM images acquired simultaneously (kcps= kilo-counts per second). Images are recorded pixel by pixel by scanning the sample under the tip from left to right and top to bottom (laser power at the uncoated tip apex: 120 $\mu$W; integration time: 80 ms; scanner speed: 1 $\mu$m s$^{-1}$; image sizes: 64 $\times$ 64 pixel$^2$). The line cut (horizontal solid line on the topography) gives the nanodiamond size (insert) at 20 nm in the present case. The arrow in (b) marks the tip position where the nanodiamond has been embarked by the scanning tip. (c) Normalized second-order time-intensity correlation function $g^{(2)}(t)$ for the functionalized tip giving evidence for single NV-center occupancy in the functionalizing ND. The red curve is an exponential fit (see Appendix B). (d) Photoluminescence spectrum of a functionalized tip (integration time: 180 s). The small peak at 575 nm (indicated by a black arrow) is the zero-phonon line of the neutral NV center.}
\end{center}
\end{figure}
Fluorescent NDs used here are produced in the same way as in our previous studies~\cite{26}. They are obtained from type Ib synthetic diamond powder with particle size centered around 25 nm and nitrogen content of 100 ppm. This powder is proton-irradiated (2.4 MeV, dose 5.1016 $H^+/cm^2$) to generate vacancies in the diamond lattice. Thermal annealing is used to activate migration of vacancies towards nitrogen impurities, thereby efficiently forming NV centers. Acid treatment and washing with water lead to a stable aqueous suspension. Intense ultra-sonification disperses the particles at their primary size and results in a very stable colloidal suspension. A 10 $\mu$l droplet of this suspension is spin-cast on a fused silica microscope cover-slip for the purpose of optical experiments. Apart from careful washing, this cover slip did not receive any particular treatment.\\
We now describe the all-optical \emph{modus operandi} that we have engineered to trap in a controlled way a well-selected single ND at the optical tip apex. The uncoated optical tip (see Appendix A) is covered with a thin layer of poly-l-lysine (molecular weight: 30000 - 70000 u), a polymer which has the property of homogenously covering the tip, including the apex (radius of curvature below 30 nm), as checked by scanning-electron microscopy. In addition, poly-l-lysine is positively charged. This facilitates electrostatic attraction of the NDs which bear negatively charged carboxylic groups on their surface due to the acid treatment. This polymer-covered tip is glued on one prong of a tuning-fork~\cite{31} for shear-force feedback and mounted in the NSOM microscope.\\
The first step of our procedure is to image the sample fluorescence to the far-field by scanning the surface under the optical tip with a very large tip-sample distance of 3 $\mu$m. This allows for selecting an interesting area with isolated NDs. In a second step, the tip is brought into the surface near-field by using shear-force regulation. A near-field fluorescence image together with a shear-force topography image are simultaneously recorded at a rather large tip-sample distance of about 50 nm (usual cruise altitudes for NSOM imaging are between 20 and 30 nm) in order to identify an isolated ND with small size and a fluorescence level among the lowest-intensity spots detected in the entire scanned area. This last point is taken as a hint that this very ND presumably hosts a single color-center. Although our setup is able to check this essential point in situ by photon-correlation counting~\cite{26}, we found it more convenient to control the single NV-occupancy after grafting of the ND.\\
\begin{figure}[h]
\begin{center}
\includegraphics[width=10cm]{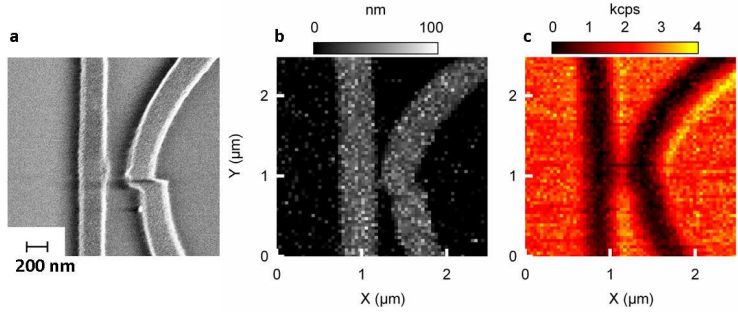}
\caption{(a) Scanning-electron micrograph of chromium structures patterned on a fused silica cover slip. (b) Numerically flattened topography of the same region. (c) Fluorescence NSOM image acquired simultanously with the scanning single-photon tip of Fig. 2 (optical power at 488 nm at the uncoated tip apex: 120 $\mu$W, integration time: 100 ms, scan height: h$\leq$ 30 nm, scanner speed: 1 $\mu$m s$^{-1}$, image size: 64 $\times$ 64 pixel$^2$). Here, the collected light is restricted by optical filtering to the emission band of the single NV center grafted on the optical tip. }
\end{center}
\end{figure}
Then, the next step is the ND attachment. This is accomplished "manually" during scanning by strengthening the shear-force feedback~\cite{31} so as to approach to the surface vertically at a distance around 30 nm at the very moment where the optical tip is facing the desired ND. This shear-force strengthening is maintained for typically two scanning lines and then released so as to bring the now functionalized tip back to its initial altitude of 50 nm. Figs. 2a,b show an example of such a trapping event: It can be seen that the ND trapping manifests itself as a sudden persistent increase in the optical signal. This increase amounts to the emission level of the ND prior to its attachment. Indeed, once the ND has been stuck at the tip apex, the tip not only transmits the excitation laser light, but also produces a background signal due to the embarked ND, irrespective of its position above the scanned surface. It is worth noting in Fig. 2b that the shear-force feedback has been forced (horizontal dashed line on the fluorescence image) after having made sure that the ND height - 20 nm in the present case - has been measured correctly in the topographic image (full line in Fig. 2a). After the attachment process, the image acquisition is completed and no additional ND is grafted to the tip thanks to the rather large tip-to-surface distance of around 50 nm. This is the reason why the 40 nm in-height ND cluster (possibly made of 3 NDs) seen in the lower left quarter in Fig. 2b is not trapped by the scanning tip. It is worth noting that an accidental fishing of an additional fluorescent ND would immediately translate into an increased fluorescence background emanating from the tip: this provides us with a "safety procedure" for ensuring that such an accidental can actually be detected.\\ Now, the functionalized tip needs further optical characterization since the embarked ND was elected from guesses that it should host a single color-center. To check this important point, we carried out photon-correlation measurements and spectrum acquisition of the functionalized tip after having moved the tip far above the surface (distance of 10 $\mu$m), laterally displaced the sample to a ND-free region, and focused the collection objective onto the probe apex. Fig. 2c shows the second-order time-intensity correlation function $g^{(2)}(t)$ (see Appendix B) for the functionalized tip obtained after subtracting the random coincidences caused by the background light~\cite{26,32}. The correlation function exhibits an anti-bunching gap at zero delay with $g^{(2)}(0)$ going far below 0.5  (i.e. $g^{(2)}(0)\simeq 0.1$). This unambiguously confirms that a single NV color-center, acting as a single photon nano-source~\cite{24}, has been attached at the tip apex. This NV center is an uncharged one as additionally revealed by the optical spectrum of Fig. 2d which exhibits the characteristic zero-phonon line of the neutral NV center at 575 nm~\cite{30}.\\
Our method is highly reproducible and reliable. We have repeatedly functionalized tips on-demand with a desired number of well-selected NDs in addition to the single ND case described above. For example, as a demonstration of the flexibility of our method, we have been able to successively seize five NDs onto a tip apex, to free all of them at once by knocking the tip on the surface, and to fish them back one by one. In addition, in contrast with previous attempts with polymer-free tips which released the embarked nanocrystals within an hour~\cite{33}, the ND remains attached at the apex for days so that the functionalized tip can not only be fully characterized, but can also be used for subsequent experiments or imaging. An example of such imaging is shown in Fig. 3. Here, we used the functionalized tip of Fig. 2 and a test sample made of 250 nm wide and 40 nm thick chromium lines and parabola that have been lithographically patterned on a fused silica cover slip. For this proof-of-principle experiment, chromium has been preferred to gold or silver to avoid undesired fluorescence from the metal. The collection light was spectrally restricted by proper filtering to the NV emission and the scanning height was set at 20 - 30 nm. As can be seen from Fig. 3, details in the structures are seen in all three images. In particular, the optical image clearly reveals the metallic structures as non-transmitting dark lines with a good contrast although it is recorded with the fluorescence light emitted by a single NV center only. The distinctive bright decorations seen on each side of the chromium structure are possibly due to the finite optical reflectivity of chromium or to modifications of the NV-center dynamical properties (e.g. change in the excited state lifetime) when the tip probe approaches the nanostructure edges~\cite{34}. Such effects have been recently reported on nanodiamond emitters located in the vicinity of metal nanostructures~\cite{6}. More experimental work at shorter tip-surface distances is required to sustain this view.\\
An interesting feature of the scanning single-photon near-field source is the spatial resolution that it can potentially offer. The chromium parabola in Fig. 3 has been patterned in such a way as to offer a variable gap with the adjacent line. Our aim was at inferring a spatial resolution for our setup from its ability of resolving two adjacent similar objects, in agreement with the basic definition of a resolving power, rather than from the lateral extension of the rise in the optical signal. In addition, for this particular line-parabola doublet shown in Fig. 3, a lithography failure brought incidentally the minimum gap to approximately 120 nm (Fig. 3a). As seen in Fig. 3c, this 120 nm gap is resolved in the optical image. This indicates that the spatial resolution is at least in this range, i.e, much better than with the initial uncoated tip which offers resolutions limited to about 400 nm~\cite{26}.\\ Moreover, the near-field optical probe reported here acts as a genuine scanning point-like dipole emitter. This is in contrast with metal-coated aperture tips which bear a polarization-dependent annular charge density around the apex~\cite{35}. As a consequence, such aperture tips exhibit a double-lobe distribution of the electromagnetic field at the apex that artificially duplicates the imaged objects so that the resolution they offer is ultimately limited by the free aperture diameter (50-100 nm) at the tip apex~\cite{35,36,37,38,39}, not by the scanning height (see Appendix C). Point-like dipolar emitters do not exhibit such a split-field distribution, so that their potential resolution should thus ultimately depend on the scanning height h only~\cite{12}. Interestingly, the ND-based active tip introduced here could fully exploit this potentiality because the NV quantum emitter is hosted by a matrix of a genuine nanometer extension.  This stresses the key role played by height control in future developments. In the present proof-of-principle experiments, we have set a safe lower bound to tip-surface distance at approximately 20-30 nm to avoid too strong friction forces to be applied to the tip apex~\cite{31}, thereby preventing a too rapid release of the 20-nm sized illuminating ND.  We were then able to use our functionalized tips for several days for image acquisition or other measurements.\\
The potential of our scanning single-photon point-like source goes well beyond high-resolution optical imaging. Indeed, an important point to emphasize is that the optical image of Fig. 3c is acquired with a single-photon source used for illumination so that only one photon is interacting with the nanostructure at any time, even in the continuous-wave regime of the laser illumination~\cite{40} (see Appendix D). This in turn means that the imaging procedure reported here can be viewed as a scanning single-photon near-field microscopy working at RT (note, however, that the nanodiamond emits an average of $10^7$ photons per pixel due to the integration time of 100 ms).  It can be anticipated that such a microscopy will offer interesting new prospects to quantum optics at the nano- and micrometer scales. First, it would be possible to study in a local and controllable way the influence of the optical environment on the emission properties of a quantum system, i.e., the NV center. In particular, the complex influence of the environment on the optical local density-of-states~\cite{34,41} (optical LDOS) could be analyzed. Other perspectives concern the strong coupling regime and the interaction between NV centers and a second quantum emitter (e.g. a colloidal quantum dot~\cite{17}) through fluorescence resonance energy transfer (FRET). Such possibilities are specific of our scanning single photon source and could not be envisaged with usual aperture NSOM probes (even coupled to an external single photon source) due to the double-lobe distribution of the optical near-field generated by the aperture~\cite{35,36,37,38,39}.
\section{Conclusion}
In summary, we have presented a simple and reproducible method for functionalizing optical tips with an individual ND hosting a single NV color-center. Our all-optical method allows each step of the tip functionalization to be carried out in a single setup working in a flexible ambient environment. This includes the selection of the object, its grafting onto the tip and its subsequent on-demand excitation by simply injecting light into the substrate optical fiber. In the latter case, such a tip realizes a genuine scanning single-photon source operating at RT that is very photostable and useable on a long-term for various experiments. We anticipate that such ND-functionalized optical tips and its related scanning single-photon microscopy will be of interest to many applications where an accurate positioning of a quantum object is essential~\cite{42,43} such as, for example, ultra-high resolution and sensitivity magnetometry~\cite{8,9,10}. Of particular interest too will be to probe the ultimate optical resolution actually offered by point-like emitters~\cite{12}. Other applications include mapping of the optical LDOS of surfaces~\cite{34,41,44} and complex plasmonic resonators~\cite{5}, local fluorescence lifetime imaging~\cite{45}, nanoscale coupling of single emitters to optical~\cite{1,2,46} or mechanical~\cite{3} resonators, and quantum optics with surface plasmons through accurate tip-launching~\cite{47} of single plasmons, i.e., quantum plasmonics~\cite{5,6,7}.
\section{Appendix A: Near-field probe fabrication}

The tapered fiber tips are prepared using the tube-etching method~\cite{48} which works as follows: First, the pure-silica core single mode optical fiber (fiber S405, Thorlabs) with its plastic cladding is dipped into a HF solution bath. The chemical etching of silica, which lasts for several 10 minutes, is assisted by HF convection in the region separating the plastic cladding from the silica part. The residual plastic cladding is finally removed using acetone. The result is a conical fiber probe with a full apex angle $\alpha\simeq 16^{\circ}$.  The final curvature radius of the fiber tip apex is typically 30 nm.\\ In order to graft a diamond  nanocrystal (functionalized  with COOH carboxylic groups) on the fiber probe we dip the tip into a positively charged poly-l-lysine solution (molecular weight 30000 -70000 $u$) for ten minutes. The tip is then retracted from the bath at the velocity 100 $\mu/s$. This polymer is thus used \emph{in situ} during the NSOM experiment to fix a nanodiamond as explained in section  2.

\section{Appendix B: Second-order correlation function [Figure 2c] }

The color center considered in the article  is an electrically uncharged Nitrogen Vacancy (NV) defect in diamond (i.e. NV$^0$). For the present purpose, the energy level structure of NV$^0$ is well described  by a two-level system  [$|e\rangle$,$|g\rangle$] completed by a metastable state $|m\rangle$~\cite{23,24,32}. Optical transitions occur between the ground  $|g\rangle$ and the excited $|e\rangle$ states and $|m\rangle$ plays the role of an optical dark state with very low probability  $k_{mg}$ to relax to $|g\rangle$. In the following we will however ignore the dark state and we will treat the system as an ideal two-level oscillator. Neglecting quantum coherence~\cite{joos} (this is justified since at room temperature non radiative transitions between energy vibration sublevels break the quantum coherence) between the different energy levels the dynamics of the NV color center can be solved using the rate equations:
\begin{eqnarray}
\frac{d}{d\tau}\left(\begin{array}{c}
p_e\\ p_g
\end{array}\right)=\left(\begin{array}{cc}
-k_{eg} & k_{ge}\\ k_{eg} & -k_{ge}
\end{array}\right)\cdot\left(\begin{array}{c}
\sigma_e\\ \sigma_g
\end{array}\right),
\end{eqnarray}
where $p_i$ are the population of the $i$ level  ($p_e+p_g=1$) and  $k_{ij}$ the transition rates~\cite{loudon}. The initial conditions, corresponding to the system in its ground state at time $\tau=0$, are written $p_e=0$, $p_g=1$ and lead, after solving the differential equation 1, to the second order correlation function~\cite{loudon}:
\begin{equation}
g^{(2)}(\tau)=\frac{p_e(\tau)}{p_e(+\infty)}=1-e^{-(k_{eg}+k_{ge})\tau}.
\end{equation} At $t=0$ equation 2 fulfills the condition $g^{(2)}(0)=0$, i.e. non classical antibunching, which characterizes the one by one single photon emission process. \\
Importantly, due to the presence of  random coincidences caused by the background light, the second order coincidence function $C_N(\tau)=\frac{\langle N(t)N(t+\tau) \rangle}{(\langle N(t)\rangle)^2}$ ($N$ is the single particle rate in  count per seconds (cps)), deduced from the photon counts recorded with the Hanbury Brown and Twiss correlator, see Figure 1, contains an additional contribution to $g^{(2)}(\tau)$. We remind that the normalized correlation $C_N(\tau)$ is connected to the histogramm of coincidences $c(t)$ given by the correlator by : $C_N(\tau)=c(t)/(\langle N_1\rangle\langle N_2\rangle W T)$ where $\langle N_1\rangle\simeq\langle N_2\rangle=9$ kcps is the single photon detection rate on the APDs 1 and 2, $W=512$ ps the time bin of the coincidence histogramm, and $T=1000$ s the total integration time. We thus have the total correlation function
\begin{equation}
C_N(\tau)=g^{(2)}(\tau)\cdot\rho^2+1-\rho^2,
\end{equation} where $\rho=\langle S\rangle/\langle(S+B)\rangle$ is the signal to signal plus noise ratio~\cite{23,24,32,33} (assuming a Poisson statistics for the noise).
In order to renormalize the correlation function one must thus know the value of $\rho$. In our experiment the noise due to the substrate  and  optical fiber fluorescence can be determined before attaching the nanodiamond onto the tip apex. This experimentally leads to a value $\rho_1\simeq 0.8$ and thus to a corrected  $\widetilde{C_N}(\tau)$:
\begin{equation}
\widetilde{C_N}(\tau)=\frac{C_N(\tau)-(1-\rho_1^2)}{\rho_1^2},
\end{equation} This should, ideally,  equal the signal correlation function $g^{(2)}(\tau)$ given by equation 2 if $\rho_1=\rho$. However, because the nanodiamond crystal itself fluoresces we have  $\rho_1>\rho$ and we obtain a small disagreement between  $\widetilde{C_N}(\tau)$ and $g^{(2)}(\tau)$ given by :
\begin{equation}
\widetilde{C_N}(\tau)=g^{(2)}(\tau)\frac{\rho^2}{\rho_1^2}+1-\frac{\rho^2}{\rho_1^2}.
\end{equation}
In particular we have $\widetilde{C_N}(0)\neq 0$ which explains the residual correlation observed at $\tau=0$ shown in Fig.2.\\ In order to fit and reproduce the experimental data we used $\frac{\rho^2}{\rho_1^2}=0.9$ and $(k_{eg}+k_{ge})^{-1}=9$ ns which give the red curve in Fig.2.  For these values  we obtain $\widetilde{C_N}(0)=0.1$ in agreement with previous observations~\cite{32}. Remark that the rate $k_{eg}+k_{ge}$ is the sum of the radiative $k_{eg}$ and pumping $k_{ge}$ rates.  It is therefore not possible to simply identify  $(k_{eg}+k_{ge})^{-1}$ with the NV color center lifetime $k_{eg}^{-1}$. The pumping rate is obtained using the condition $\langle N\rangle=\eta\cdot k_{eg}p_e(+\infty)=\eta k_{eg}k_{ge}/(k_{eg}+k_{ge})$ with $\langle N\rangle\simeq 9 $ kcps the average photon rate on the amplified photon detector (APD), and  $\eta$ the total collection-detection efficiency of the whole optical setup including  the APD.   Considering the properties of the different elements of our setup we estimate $\eta=1\%$. This leads to $k_{ge}^{-1}= 1$ $\mu$s  and thus to the lifetime $k_{eg}^{-1}\simeq 9 $ ns. This value agrees with those usually reported (i.e. $k_{eg}^{-1}\simeq 10-20$ ns) for the considered nanodiamonds ~\cite{7,22,23,32}. More studies have however to be done to understand precisely the quantum efficiency of our single photon source.
\section{Appendix C: Optical resolution in near field microscopy}
\begin{figure}[h]
\begin{center}
\includegraphics[width=11cm]{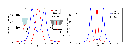}
\caption{(a) Simulation of the optical image obtained by scanning a fluorescent isotropical emitter at a constant height $h=20$ nm below the NSOM tip. The red curve  is the theoretical result obtained with a dipolar point like source (like the NV center). It is compared with the image obtained with an usual aperture NSOM, hole radius: 50 nm, (blue curve). (b) same as (a) but for $h=10$ nm.  The incident light polarization direction and the polarization dipole orientation of the active probe are both aligned with the scan direction.}
\label{truc}
\end{center}
\end{figure}
The optical resolution of aperture near-field optical microscopy is intrinsically limited by the geometrical diameter of the hole located at the apex of the probe.  To simulate the optical field generated by such NSOM tips we here use the method developed in refs.~\cite{35,39}.  More precisely, we assume that the electric near-field in the vicinity of the aperture is equivalent to the one created by a linear distribution of charge located along the aperture rim  with an azimuthal dependence
\begin{equation}
\sigma(\phi)=\sigma_0\cdot\cos{(\phi)},
\end{equation}
where $\phi$ is the azimuthal angle in the aperture plane relatively to the linear polarization of the incident light~\cite{35,39}.  In order to mimic what happens during a typical scan of a sample below the NSOM tip we calculate the field generated by the charge distribution given by Eq.~6  on a fluorescent emitter located in front of the tip at a constant height $h$. We show in Fig.~\ref{truc} the power radiated during the scan by such an emitter that we suppose isotropical for simplicity (such assumption is justified if one uses for example a fluorescent latex nanosphere as typical emitter~\cite{39}). The two-lobe signal is a direct signature of the ring-like charge distribution as confirmed by earlier experiments~\cite{39}. We compare this result with the prediction given, in the same conditions, by a simple point-like dipole which mimics the optical behavior of our NV-center based probe (see Fig.~\ref{truc}). The differences are very significant due to the complexity of the optical near-field generated by the aperture NSOM compared to a single-dipole electric field. For this reason, the spatial resolution of our single photon near-field microscopy is ultimately only limited by the height $h$ as can be seen by comparing Fig.~\ref{truc} a and b. This is clearly not the case with an aperture NSOM tip.
\section{Appendix D : Single photon near-field microscopy}
The typical time between two successive photon emissions is given by the excited-state lifetime $\tau$. During this lifetime a photon propagates over a maximum distance of $c\tau=3- 6$ m ($c=$ light celerity) which is by orders of magnitude larger than any scale of the scanned image. Going back to the correlation function in Figure 2c, it is thus clear that at the time $T=L/c$, where $L$ is a characteristic propagation length of the photon in the imaged structure (in the nanometer or micrometer range) $g^{(2)}(T)$ is practically equal to zero. Since $g^{(2)}(T)$ represents the conditional probability to detect a second photon at time $t+T$ provided that a first one has already been recorded at time t, normalized to the single photon detection probability at time t~\cite{loudon}, the small value of $g^{(2)}(T)$ confirms that only one photon at once interacts with the structure.


\begin{references}
\bibitem{1} Y.-S. Park, A. K. Cook, and H. Wang,  ``Cavity QED with diamond nanocrystals and silica microspheres,'' Nano Lett. \textbf{6}, 2075-2079 (2006).

\bibitem{2} S. Schietinger, T. Schr\"{o}der, and O. Benson,  ``One-by-one coupling of single defect centers in nanodiamonds to high-Q modes of an optical microresonator,'' Nano Lett. \textbf{8}, 3911-3915 (2008).

\bibitem{3} J. D. Thompson,  et al. ``Strong dispersive coupling of a high-finesse cavity to a micromechanical membrane,'' Nature \textbf{452}, 72-75 (2008).

\bibitem{4} A. Cleland,  ``Optomechanics: Photons refrigerating phonons ,'' Nature Phys \textbf{5}, 458-460 (2009).

\bibitem{5} N. Verellen,  et al. ``Fano resonances in individual coherent plasmonic nanocavities,'' Nano Lett. \textbf{9}, 1663-1667 (2009).

\bibitem{6} S. Schietinger, M. Barth, T. Aichele, and O. Benson,  ``Plasmon-enhanced single photon emission from a nanoassembled metal-diamond hybrid structure at room temperature,'' Nano Lett. \textbf{9}, 1694-1698 (2009).

\bibitem{7} R. Koselov,  et al. ``Wave-particle duality of single surface plasmon polaritons,'' Nature Phys. \textbf{5}, 470-474 (2009).

\bibitem{8} C. L. Degen,  ``Scanning magnetic field microscope with a diamond single-spin sensor,'' Appl. Phys. Lett. \textbf{92}, 243111 (2008).

\bibitem{9} J. R. Maze,  et al.,  ``Nanoscale magnetic sensing with an individual electronic spin in diamond,'' Nature \textbf{455}, 644-647 (2008).
\bibitem{10} G. Balasubramanian, et al., ``Nanoscale imaging magnetometry with diamond spins under ambient conditions,'' Nature \textbf{455}, 648-651 (2008).

\bibitem{11} W. E. Moerner, and M. Orrit,``Illuminating single molecules in condensed matter,'' Science \textbf{283}, 1670-1676 (1999).

\bibitem{12} J. Michaelis, C. Hettich, J. Mlynek,  and V. Sandoghdar,  ``Optical microscopy using a single-molecule light source,'' Nature \textbf{405}, 325-328 (2000).

\bibitem{13} P. Michler, A. Imamoglu, M. D. Mason, P. J. Carson, G. F. Strouse,  and S. K. Buratto,  ``Quantum correlation among photons from a single quantum dot at room temperature,'' Nature \textbf{406}, 968-970 (2000).

\bibitem{14} B. Mahler, P. Spinicelli, S. Buil, X. Quelin, J.-P. Hermier, and B. Dubertret,  ``Towards non-blinking colloidal quantum dots,'' Nature Mat. \textbf{7}, 659-664 (2008).

\bibitem{15} K. T. Shimizu, R. G. Neuhauser, C. A. Leatherdale, S. A. Empedocles, W. K.Woo,  and  M. G. Bawendi, ``Blinking statistics in single semiconductor nanocrystal quantum dots,''  Phys. Rev. B \textbf{63}, 205316 (2001).

\bibitem{16} N. Chevalier,  et al., ``CdSe single-nanoparticle based active tips for near-field optical microscopy,'' Nanotechnology \textbf{16}, 613-618 (2005).

\bibitem{17} X. Wang,  et al.,  ``Non-blinking semiconductor nanocrystals,'' Nature \textbf{459}, 686-689 (2009).

\bibitem{18} Y. Sonnefraud,  et al. , ``Near-field optical imaging with a CdSe single nanocrystal-based active tip,'' Opt. Express \textbf{14}, 10596-10602 (2006).

\bibitem{19} A. Tribu,  et al., `` A high-temperature single-photon source from nanowire quantum dots,''  Nano Lett. \textbf{8}, 4326-4329 (2008).

\bibitem{20}  D. Giaume, et al., ``Organic functionalization of luminescent oxide nanoparticles toward their application as biological probes,'' Langmuir \textbf{24}, 11018-11026 (2008).

\bibitem{21} A. Cuche,  et al., `` Fluorescent oxide nanoparticles adapted to active tips for near-field optics,'' Nanotechnology \textbf{20}, 015603 (2009).

\bibitem{22} A. Gruber, A. Dr\"{a}benstedt, C. Tietz, L. Fleury, J. Wrachtrup,  and C. Von Borczyskowski, ``Scanning confocal optical microscopy and magnetic resonance on single defect centers,'' Science \textbf{276}, 2012-2014 (1997).

\bibitem{23} A. Beveratos, R. Brouri, T. Gacoin, J.-P. Poizat,  and P. Grangier,  ``Nonclassical radiation from diamond nanocrystals,'' Phys. Rev. A \textbf{64}, 061802 (2001).

\bibitem{24} A. Beveratos, S. K\"{u}hn, R. Brouri, T. Gacoin, J.-P. Poizat, and P. Grangier, ``Room temperature stable single-photon source,'' Eur. Phys. J. D \textbf{18}, 191-196 (2002).

\bibitem{25} Y.-R. Chang,  et al., ``Mass production and dynamic imaging of fluorescent nanodiamonds,'' Nature Nanotechnol \textbf{3}, 284-288 (2008).

\bibitem{26} Y. Sonnefraud,  et al., ``Diamond nanocrystals hosting single nitrogen-vacancy color centers sorted by photon-correlation near-field microscopy,'' Opt. Lett. \textbf{33}, 611-613 (2008).

\bibitem{27} J.-P. Boudou,  et al., ``High yield fabrication of fluorescent nanodiamonds,'' Nanotechnology \textbf{20}, 235602 (2009).

\bibitem{28} B. R. Smith,  et al.,`` Five-nanometer diamond with luminescent nitrogen-vacancy defect centers,'' Small \textbf{5}, 1649-1653 (2009).

\bibitem{29} S. K\"{u}hn, C. Hettich, C. Schmitt, J.-P. Poizat, and V. Sandoghdar,  ``Diamond colour centres as a nanoscopic light source for scanning near-field optical microscopy,'' J. Microsc. \textbf{202}, 2-6 (2001).
\bibitem{30} Y. Dumeige, F. Treussart, R. Alléaume, T. Gacoin,  J.-F. Roch,  and P. Grangier,  ``Photo-induced creation of nitrogen-related color centers in diamond nanocrystals under femtosecond illumination,'' J. Lumin. \textbf{109}, 61-67 (2004).

\bibitem{31} K. Karrai,   and R. D. Grober,  ``Piezoelectric tip-sample distance control for near field optical microscopes,'' Appl. Phys. Lett. \textbf{66}, 1842-1844 (1995).

\bibitem{32} R. Brouri, A. Beveratos , J. P. Poizat,  and P. Grangier,  ``Photon antibunching in the fluorescence of individual color centers in diamond,'' Opt. Lett. \textbf{25}, 1294-1296 (2000).

\bibitem{33}  A. Cuche, et al., `` Diamond nanoparticles as photoluminescent nanoprobes for biology and near-field optics,''  J. Lumin., doi: 10.1016/j.jlumin.2009.04.089 (2009).

\bibitem{34} C. Girard, O. J. F. Martin, G. Leveque,  G. Colas des Francs, A.  Dereux,  ``Generalized bloch equations for optical interactions in confined geometries,'' Chem. Phys. Lett. \textbf{404}, 44-48 (2005).

\bibitem{35} A. Drezet, S. Huant,  and J. C. Woehl,  ``In situ characterization of optical tips using fluorescent nanobeads,'' J. Lumin.  \textbf{107}, 176-181 (2004).

\bibitem{36} E. Betzig,  and R. J. Chichester,  ``Single molecules observed by near-field scanning optical microscopy,'' Science \textbf{262}, 1422-1425 (1993).

\bibitem{37} J. K.  Trautman, J. J. Macklin, L. E. Brus,  and E. Betzig,``  Near-field spectroscopy of single molecules at room temperature,'' Nature \textbf{369}, 40-42 (1994).

\bibitem{38} N. F. van Hulst, J.-A. Veerman, M. F. Garcia-Parajo,  and L. Kuipers,``  Analysis of individual (macro)molecules and proteins using near-field optics,'' J. Chem. Phys.  \textbf{112}, 7799-7810 (2000).

\bibitem{39} A. Drezet, M. J. Nasse, S. Huant,  and J. C.Woehl,  ``The optical near-field of an aperture tip,'' Europhys. Lett. \textbf{66}, 41-47 (2004).

\bibitem{40} L. Mandel,  and E. Wolf,  \emph{Optical coherence and quantum optics} (Cambridge Univ. Press, Cambridge, 1995).

\bibitem{41}  G.  Colas des Francs, et al., ``Optical analogy to electronic quantum corrals,'' Phys. Rev. Lett. \textbf{86}, 4950-4953 (2001).

\bibitem{42}T. Van der Sar,  et al.,  ``Nanopositioning of a diamond nanocrystal containing a single nitrogen-vacancy defect center,'' Appl. Phys. Lett. \textbf{94}, 173104 (2009).

\bibitem{43} E. Ampem-Lassen,  et al., `` Nano-manipulation of diamond-based single photon sources,'' Opt. Express \textbf{17}, 11287-11293 (2009).

\bibitem{44} A. Sundaramurthy,  P. J.Schuck, N. R.Conley, D. P.Fromm ,  G. S. Kino, and W. E.Moerner , ``Toward nanometer-scale optical photolithography:  Utilizing the near-field of bowtie optical nanoantennas,'' Nano Lett. \textbf{6}, 355-360 (2006).

\bibitem{45} J. Tisler,  et al., `` Fluorescence and spin properties of defects in single digit nanodiamonds,'' ACS Nano \textbf{3}, 1959-1965 (2009).

\bibitem{46} T. H. Taminiau, F. D. Stefani, F. B. Segerink,  and N. F. Van Hulst,  ``Optical antennas direct single-molecule emission,'' Nature Photon. \textbf{2}, 234-237 (2008).

\bibitem{47} M. Brun, A. Drezet, H. Mariette,  N. Chevalier, J. C. Woehl,   and S. Huant,  ``Remote optical addressing of single nano-object,'' Europhys. Lett. \textbf{64}, 634-640 (2003).
\bibitem{48}
R. St\"{o}ckle, et al.,  Appl.~Phys.~Lett. \textbf{75}, 160-162 (1999).
\bibitem{joos}
E. Joos, H.~D.Zeeh, C. Kiefer, D. Giulini, J. Kupsch, I.-O. Stamatescu,  \emph{Decoherence and the Appearance of a classical World in Quantum Theory, 2$^{nd}$ ed.} (Springer, New York, 2003).
\bibitem{loudon}
R. Loudon, \emph{The quantum theory of light} (Oxford University Press, New York, 2000).




\end{references}
\end{document}